# Micro-transfer printing of GaSb optoelectronics chips for mid-infrared silicon photonics integrated circuits

*Heidi Tuorila*[1], *Jukka Viheriälä*[1], *Yeasir Arafat*[2], *Fatih Bilge Atar*[2], *Fatima Gunning*[2], *Brian Corbett*[2], *and Mircea Guina*[1]

1) Optoelectronics Research Centre, Physics Unit, Tampere University, 33101 Tampere, Finland

2) Tyndall National Institute, University College Cork, Cork Ireland

* E-mail: heidi.tuorila@tuni.fi

**Abstract**

3D integration of GaSb-based gain chips on a silicon photonics platform using micro-transfer printing is demonstrated for the first time. The release process of GaSb coupons, and their transfer for the demonstration of hybrid GaSb/Silicon-photonics on-chip external cavity lasers is reported. A methodology to evaluate the key features of the gain chip coupons, namely the quality of the etched facets and the facet coating deposited using a wafer-level process, is introduced. The characterization provides insight into the fabrication factors limiting the performance of the gain coupons. The level of performance achieved for the transfer printing process offers a solid landmark for the development of photonics integration technology operating at the 2-3 µm wavelength range. This is instrumental for the deployment of mid-IR photonic integration technology in emerging applications related to gas and biomarker sensing.

## Introduction

Micro-transfer printing (µTP) is an emerging approach for the integration of III-V optoelectronics components, such as optical amplifiers, laser diodes and detectors with silicon photonics (SiPh) platforms[1,2]. The wavelength versatility enabled by the hybrid integration of III-V and SiPh elements is increasingly important for extending the applications from the established datacom solutions operating near 1.5 µm to sensing applications requiring operation at mid-infrared (MIR) wavelengths. For example, the wavelength region beyond 2 µm is relevant for monitoring environmental gases [3] and biomedical markers, such as the lactates, glucose, or urea [4–6]. In general, the demand for compact multipurpose sensors is rapidly growing, pushed by the expansion of sports and wearable medical devices[7]. For example, the current sport watches already offer a variety of monitoring capabilities, such as blood oxygen saturation ($SpO_2$) and temperature, yet they are still limited in ability to sense other target biomarkers. This limitation is generally linked to wavelength extension capability rather than the need to deploy more advanced sensing techniques. To this end, we note that both silicon-on-insulator (SOI) [8] and SiN [9] integration platforms exhibit low losses in the 2-3 µm wavelength range. Moreover, when it comes to the III-V components, GaSb-based gain heterostructures enable light generation functions in the 1.7-4 µm wavelength range[9,10] using Type-I quantum-wells (QWs). The use of Type-1 QWs optoelectronic devices in mid-IR PICs brings important benefits in terms of power consumption owing to low operating voltages (~1 V range) and moderate currents; this is an instrumental feature for wide-scale deployment of PIC technology in optical sensor applications, for example in wearable devices.

While there has been recent promising work showing that GaSb grown directly on Si wafers can be used for regrowth of lasers [11,12], monolithic integration is still very far from possible implementations in development scenarios targeting a full-scale photonics integrated circuit (PIC) functionality. To this end, various hybrid integration methods for the 2 µm range have been recently reported. For example, SOI-based heterogeneously integrated lasers were demonstrated at 2.1 µm [13] and 2.3 µm [14] based on InP amplifiers integrated on silicon waveguides using wafer bonding and evanescent coupling. A hybrid tunable laser was demonstrated by Wang et al. [15] using GaSb, and by Sia et al.[16] (in this case an InP-based gain chip was coupled to the SOI circuit using an end-fire approach). Moreover, a DBR laser was demonstrated by Zia et al. [17] using flip-chip bonding of a GaSb gain block on SOI.

Compared to more established approaches for hybrid integration of III-V and SiPh components, micro-transfer printing has recently emerged as an attractive method offering a good trade-off between versatility of flip-chip integration solutions of single components and the scalability offered by wafer-scale heterogeneous integration[1]. Essentially, µTP allows assembly of various types of III-V building blocks having a minimal functional size, while being able to transfer and integrate a single coupon or thousands of units in a single step. An essential step of the µTP process, which is schematically described in Figure 1, is the fabrication of the III-V transferable coupons by releasing them from their native substrate. This technique has been increasingly adopted for the integration of GaAs [2,18–25] and InP [26–38] based components on both SOI and SiN platforms, benefiting from mature fabrication processes available for these material systems.

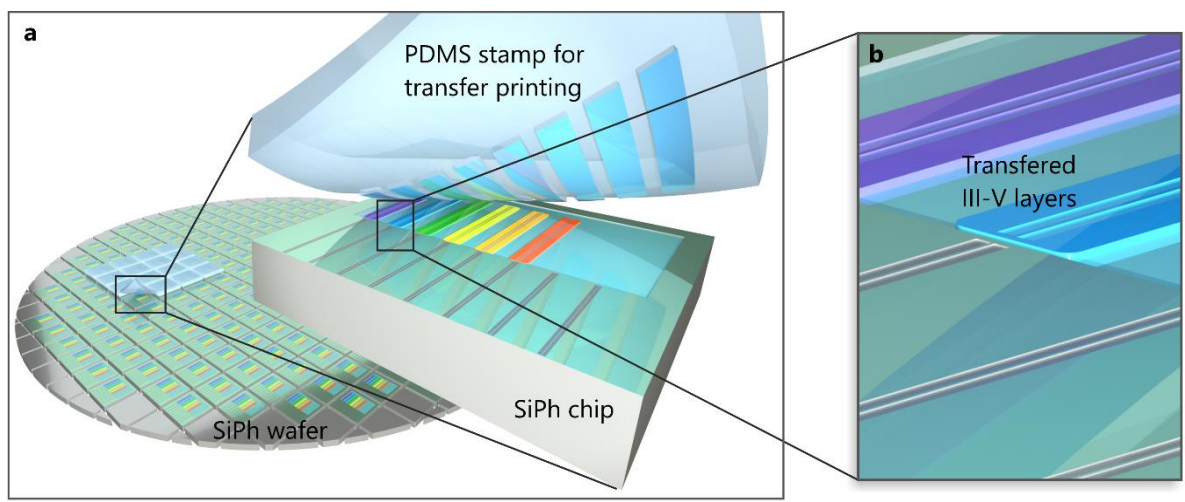

Figure 1. Transfer print concept. III-V coupon transfer on SiPh platform using a PDSM stamp.

Here we report the first demonstration of a µTP process applied for the transfer of GaSb optoelectronic components, i.e. reflective semiconductor optical amplifier (RSOA). The critical aspect providing the differentiation is the process development enabling the release of the coupons from the GaSb substrates. The performance of the µTP is validated by demonstrating continuous-wave (CW) operation of a SiPh integrated distributed Bragg reflector (DBR) laser containing the RSOA gain block µTP on a SOI circuit. To our best knowledge, no GaSb based devices involving µ-TP has been demonstrated so far, albeit some initial release etch development reported for solar cells [39]. In this respect, the schematic presented in Figure 2 illustrates an overview of the current status of the µTP developments covering different optoelectronic material systems [2,18–37,40–48]. The schematic also reveals the corresponding wavelengths covered by the different material systems and the transmission window for main SiPh integration platforms, i.e. SiN, SOI, and Ge.

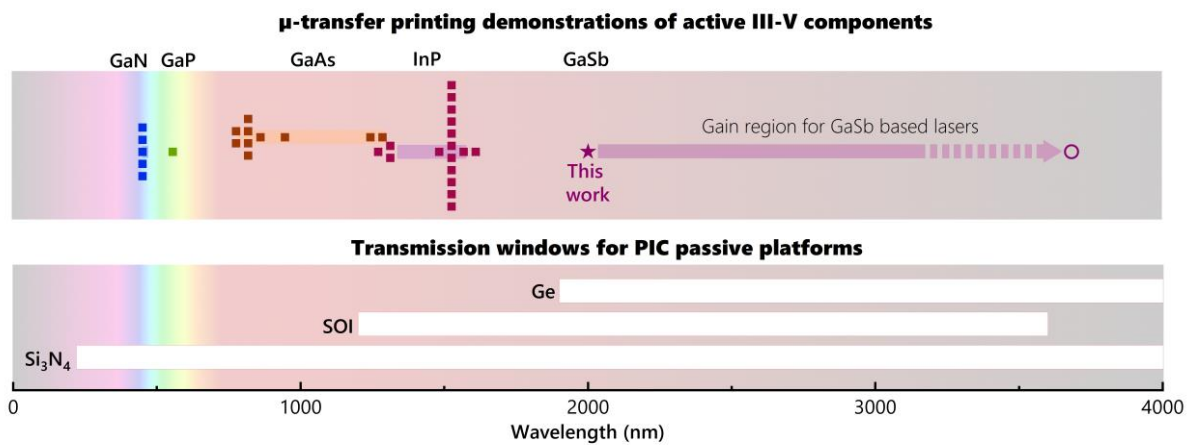

Figure 2. Overview of the published µTP device results across the demonstrated/target wavelength and the used material systems including active III-Vs and passives like Si and Ge. The data points are spread along the y-axis to display each point separately.

While the integrated lasers demonstrated in this paper operate at 2 µm wavelength range, the Type-1 quantum well GaSb gain chips allow wavelength extension up to 3.7 µm [49].

## GaSb device heterostructure

The optoelectronic heterostructure intended for transfer printing are epitaxially grow on a native III-V substrates on top of a chemically reactive sacrificial layer, i.e. a release layer, which allows their separation through wet etching. The release layer is isolated from the active structure to be transferred by a chemically inert etch stop layer. The basic steps of a generic µTP process are illustrated in Figure 3. After the fabrication of waveguides and electric contacts the first specific step is the definition of the single devices, i.e. coupons, a step that also includes the formation of etched facets acting as in-out optical interfaces (see Figure 3 a). After this, the release layer is exposed, and a supporting resist tether lateral structure is applied. The tether structure supports the active layers for the subsequent etch step in which the sacrificial release layer is removed by wet etching (see Figure 3 b). Finally, the suspended coupons are picked up with a PDMS stamp and transferred onto the silicon photonics platform where they are aligned with the passive waveguides (see Figure 3 c-d).

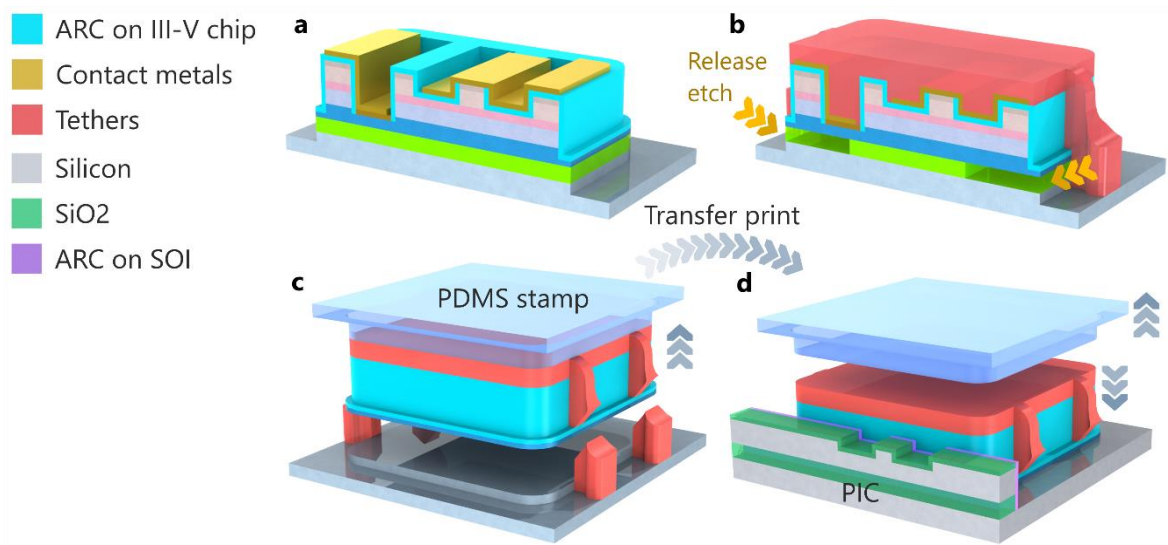

Figure 3. Micro-transfer-printing concept and process flow. a) Fabrication of device features, such as antireflection coating, insulator, and contact metal deposition followed by the exposure of the release layer; b) Application of supporting resist tethers and release etch; c) Coupon pick-up with a PDMS stamp; d) Coupon placement on silicon photonics platform and stamp removal.

An important part of this work entails the definition of a methodology used to study the functionality of transfer printed device. To this end, we introduce a systematic approach to isolate the influence of different subparts involved in the transfer print process enabling to correlate the device operation with process parameters and pinpoint the development needs. To this end, the transfer printable coupons fabricated have been grouped into several experimental series (designated as TX) corresponding to different functionalities as illustrated in Figure 4. The detailed experimental procedure to test relevant features using the series of µTP devices is described in the Methods section.

With reference to the schematic description shown in Figure 4, device types T1 and T3 were defined as reflective semiconductor optical amplifiers (RSOAs) while the device types T2 and T4 were intended to be tested as Fabry-Perot laser diodes (FP-LDs). We note that a laser diode sample offers a fast and facile procedure for assessing the performance of the transfer process given the facile assessment of the output characteristics and correlation to the waveguide and mirror loss. The transfer printable coupons are compared against standard discrete FP-LD chips (T5 to T8) to reveal how the specific features required for the implementation of the transfer printing process are impacting the device performance.

The device fabrication series were based on two epitaxial fabrication runs designated as Epi 1 and Epi 2. The samples compared as part of each series were used to analyze the following functionalities (i.e. the interplay between the fabrication features and of device performance):

1) *Definition of the n-contacts on the epi-side of the wafer (samples T3 & T4). Conventional discrete GaSb-gain chips and diode lasers have the n-contact on the backside (substrate part). This offers a larger contact area and ease of fabrication. However, in µTP the drive current is provided from the top of the chip and the chip backside is reserved for the formation of a mechanical and thermal contact with the PIC. The challenge for the n-contact formation is that the coupon is relatively narrow (~80 µm) so the area available for the n-contact is limited. This adds to the pre-existing complexity in forming ohmic contacts on n-GaSb owing to the Fermi level pinning [50].*

2) *To study the performance of the facet etching process we used the so-called T-bar waveguide geometry (see the difference between T5 and T6). Compared to standard waveguide geometry shown in T5, the T-bar structure features a safety margin between the ridge waveguide (RWG) and the etched facet. It should also be noted that we performed a separate study to optimize the etch through the epi-layers up to the etch stop layer. This etching step was used to define the optical output facet and the gain cavity; it replaces the cleaving step used to prepare standard discrete devices. Note, that for RSOAs, the front facet of the device is tilted 7° in relation to the optical axis to reduce back reflections.*

3) *Analysis of the optical coating deposition at wafer level. Specifically, the RSOA require a dielectric antireflection coating (ARC) to minimize the front facet reflections; the coating acts also as a facet protections during the wafer level fabrication steps. Moreover, a coating mimicking the reflection from a cleaved facet with a straight waveguide was used as a reference to compare the performance of the µTP FP-LD chips and standard FP-LDs. This coating is here referred as a laser diode coating (LDC). Finally, a high reflection coating (HRC) deposited at wafer level is necessary for the back facets to optimize the RSOA operation.*

4) *Analysis of the insertion losses induced by fabrication process of the etched waveguide facet. Device facets are typically approximated as lossless, partially reflecting mirrors. However, here we consider the additional losses related to the T-bar geometry, etched facets, and their coating. In our analysis these losses are separated from the concept of the mirror loss that is usually used to describe the portion of the power coupled out of the device responsible for the usable device output power. We refer to the additional loss as a "facet insertion loss" or "facet loss" in short.*

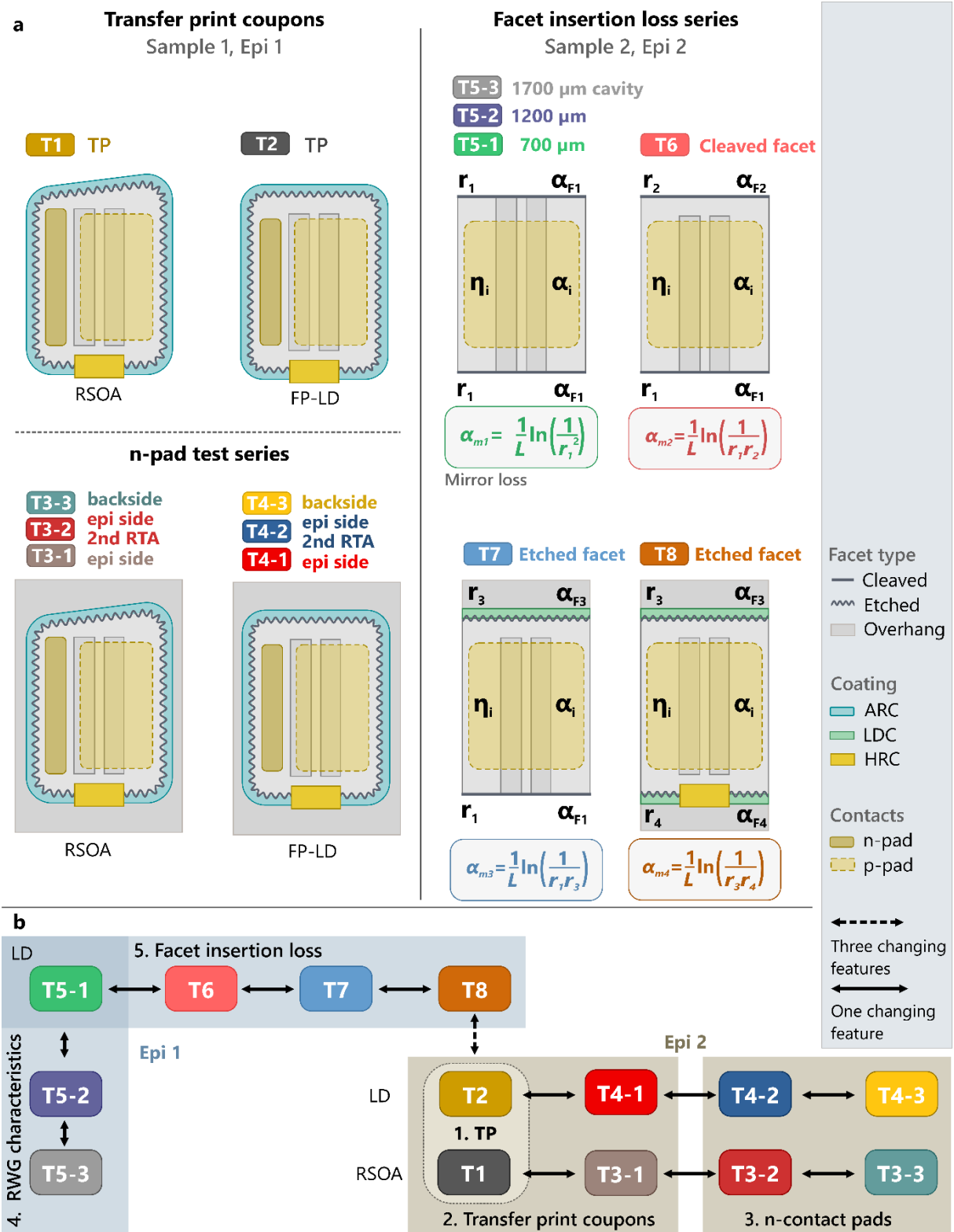

Figure 4. a) Transfer print devices and test device types. Transfer print coupon series contains devices with a structure necessary for TP. Facet loss -series consists of Fabry-Perot laser diode devices with design variants for testing various parameters (see **Supplementary information**). b) Device type chart and their relations defined in terms of the functionality studied.

## Characterization results of test series

We first analyze the results for the transfer printed devices (T1, T2). The interplay between the device structure and the process parameters was analyzed by comparing the results for the test devices (T3-T8). To start with, microscopy pictures of devices transferred onto a Si substrate and a functional SOI substrate with DBR mirrors are presented in Figure 5.

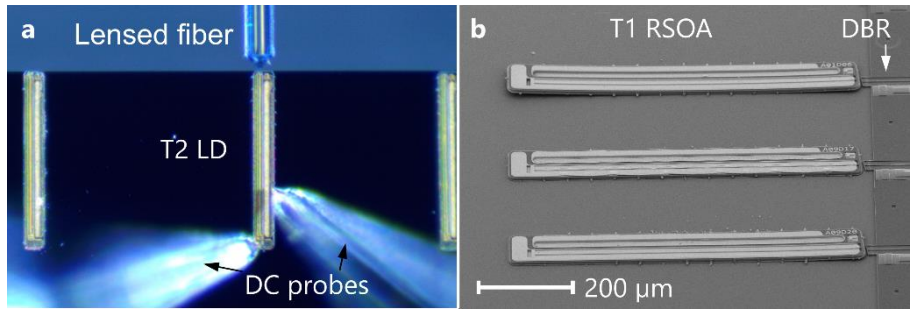

Figure 5. a) Microscope image of transfer printed coupons on plain Si substrate. The picture also displays the position of the DC probe for current injection and the alignment with an outcoupling fiber; b) SEM micrograph of transfer printed RSOA coupons on SOI platform with DBR mirrors.

The output power and device voltage were measured from each device type. The light-current-voltage (LIV) graphs for the T1 and T3 RSOA and T2 and T4-1 FP-LD devices are displayed in the Figure 6 for the comparison. The rest of the results are presented in the next section.

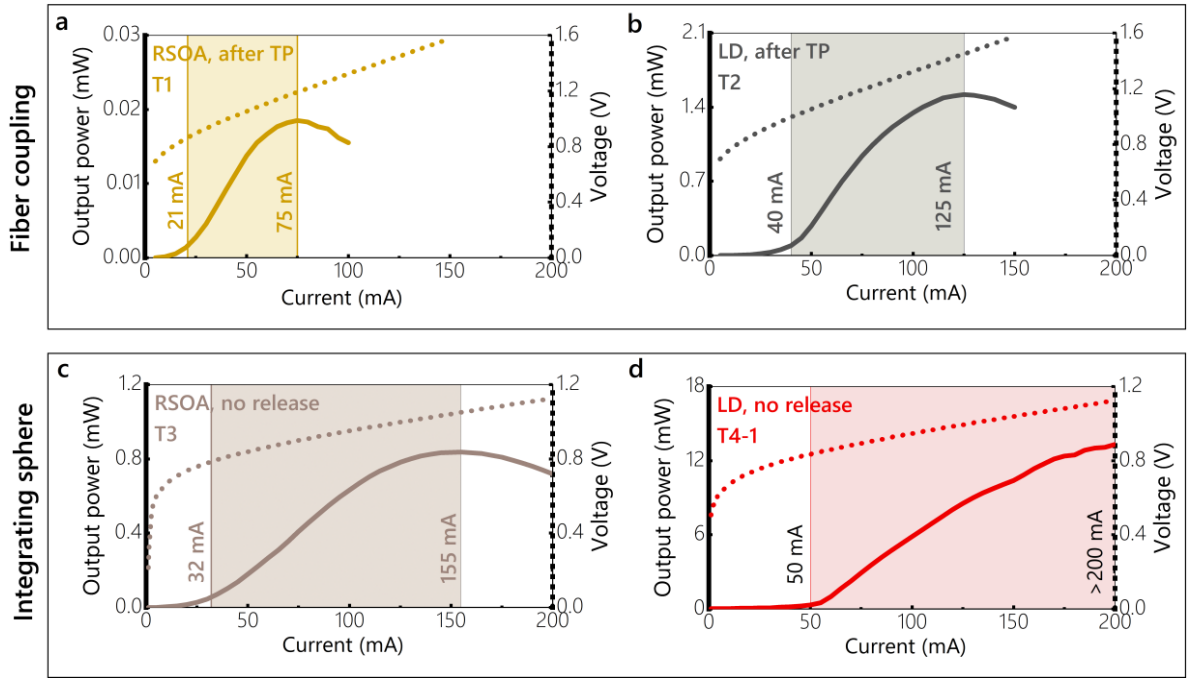

Figure 6. LIV characteristics of a) RSOA and b) FP-LD coupons after release etch and transfer printing on a plain Si wafer measured with a lensed fiber; c) RSOA and d) FP-LD coupons before release etch measured using an integrating sphere. Threshold currents and thermal rollover currents have been marked for each LI plot.

For characterization and comparison of the transfer printed T1 RSOA and T2 FP-LD, the devices on a plain Si target were biased in continuous wave with DC probes at room temperature and the light was collected with a single mode lensed fiber (Figure 5 a). The measured threshold of the transfer printed T1 and T2 devices were around 21 mA and 32 mA, respectively. The LIV graphs of transfer printed devices are shown in Figure 6 a and b. A rollover of optical power was observed due to heating at high injection current.

The test devices T3-T8 were mounted on a sub-mount and measured at room temperature in CW operation with an integrating sphere. The LIV curves from the T3 RSOA and T4-1 FP-LD without release etch are presented in Figure 6 c and d. For these test devices the threshold current is ~10 mA higher than for their TP counter parts but most importantly, the thermal rollover current is also significantly higher. The decrease in the roll-over current for TP devices can be explained by increased thermal resistance between the GaSb coupon and the Si substrate, possibly caused by the presence of the BCB layer used as a glue in this case. However, this does not explain the decrease in the threshold current.

Figure 7 displays the spectra from the RSOA devices before release etch and after transfer printing (T3 and T1). Measurements correspond to room temperature CW operation for an injection current of 150 mA for T3 and 60 mA for T1. The light was collected into a cleaved, bare fiber positioned at ~0.5 mm distance of the facet of the RSOA, and a lensed fiber positioned at a shorter distance. The spectra observed for both devices exhibit some spectral modulation, which are more pronounced in the case of T1. The measured mode spacing is 0.7 nm, corresponding to a cavity length of 700 µm. This indicates a need for improved suppression of reflections in the cavity by increasing the fabrication accuracy of the wafer-level ARC deposition.

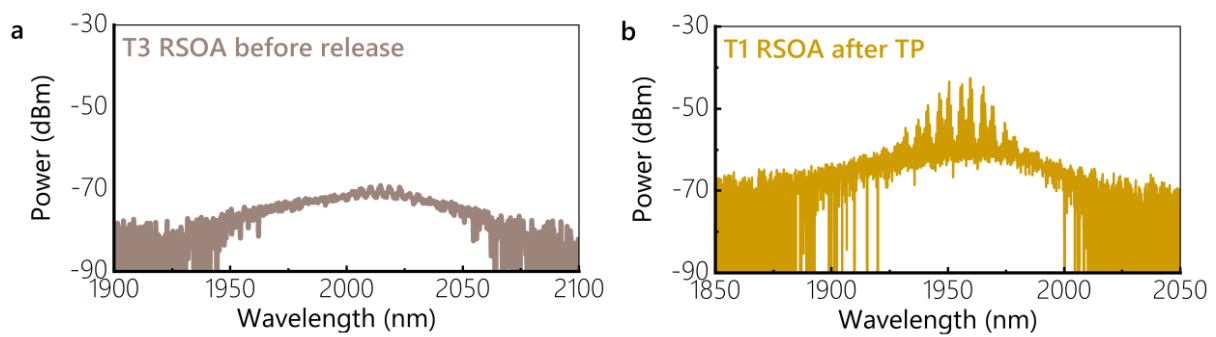

Figure 7. Emission spectra of a) T3 RSOA at 150 mA before release etch; and b) T1 RSOA at 60 mA after transfer printing.

**Performance of the integrated laser**

The spectral characteristics of the integrated laser are shown in Figure 8. To avoid thermal effects, the RSOA was driven in pulsed mode (pulse period of 10 kHz, pulse width 500 ns). The optical spectrum for operation at 90 mA reveals a single longitudinal mode (see Figure 8 a). The side mode suppression ratio (SMSR) was >14 dB. This relatively low value is attributed to a poor alignment between RSOA and the SOI waveguides resulting in low feedback, and the gap between RSOA-WG and SOI-WG. Stable DBR reflections were observed under different bias conditions of the RSOA (see Figure 8 b), revealing a good match between calculated wavelength response of the DBR at 1960 nm and the measured reflection at 1959.5 nm.

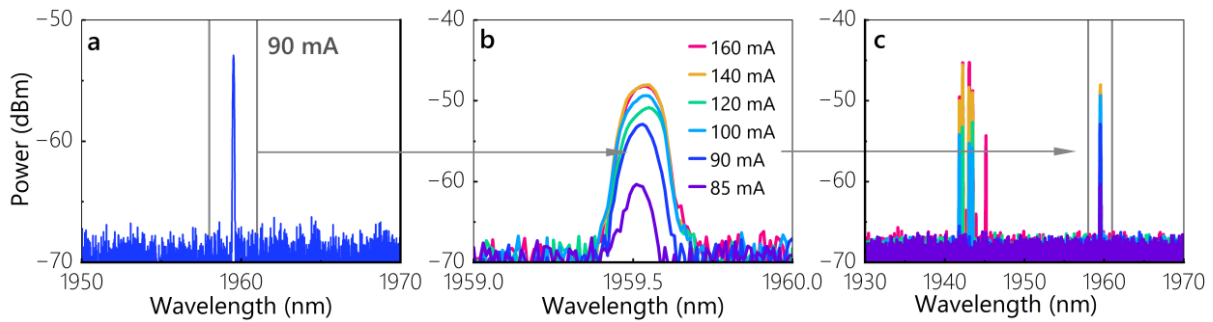

Figure 8. Emission spectra of the external cavity laser measured in pulsed operation at room temperature. a) Spectrum revealing single mode lasing at 90 mA. b) Close-up of the 1959.5 nm emission peaks corresponding to DBR reflection. c) Spectra measured for different injection currents. The side modes at shorter wavelength appear for injection currents ≥100 mA.

When operating the device at bias currents ≥100 mA, outside the nominal operation region where the thermal effects are already taking place, extra peaks were observed in the shorter wavelength region compared to the DBR reflection. The presence of these side modes most likely indicates that the ARCs at the RSOA front facet or at the SOI facet does not suppress the back reflections sufficiently.

**Analysis of light-current characteristics of the µTP devices**

To further understand the relation between the µ-TP process and the overall device operation, Figure 9 presents a comparative data set grouped by several operational characteristics of the laser diodes, i.e. threshold current densities, thermal rollover currents, and differential quantum efficiencies. The differential quantum efficiencies ($\eta_d$) are displayed only for the front facets as the ratios of output power between front and back facets vary from device to device due to different types of optical coatings used (HRC/LDC). The effect of the dissimilar coatings between the front and back facets on the distribution of the total $\eta_d$ between the facets is further discussed in the Supplementary Information. For the FP-LD the total $\eta_d$ is double the value measured from the front facet with the assumption of equal mirrors on the front and the back.

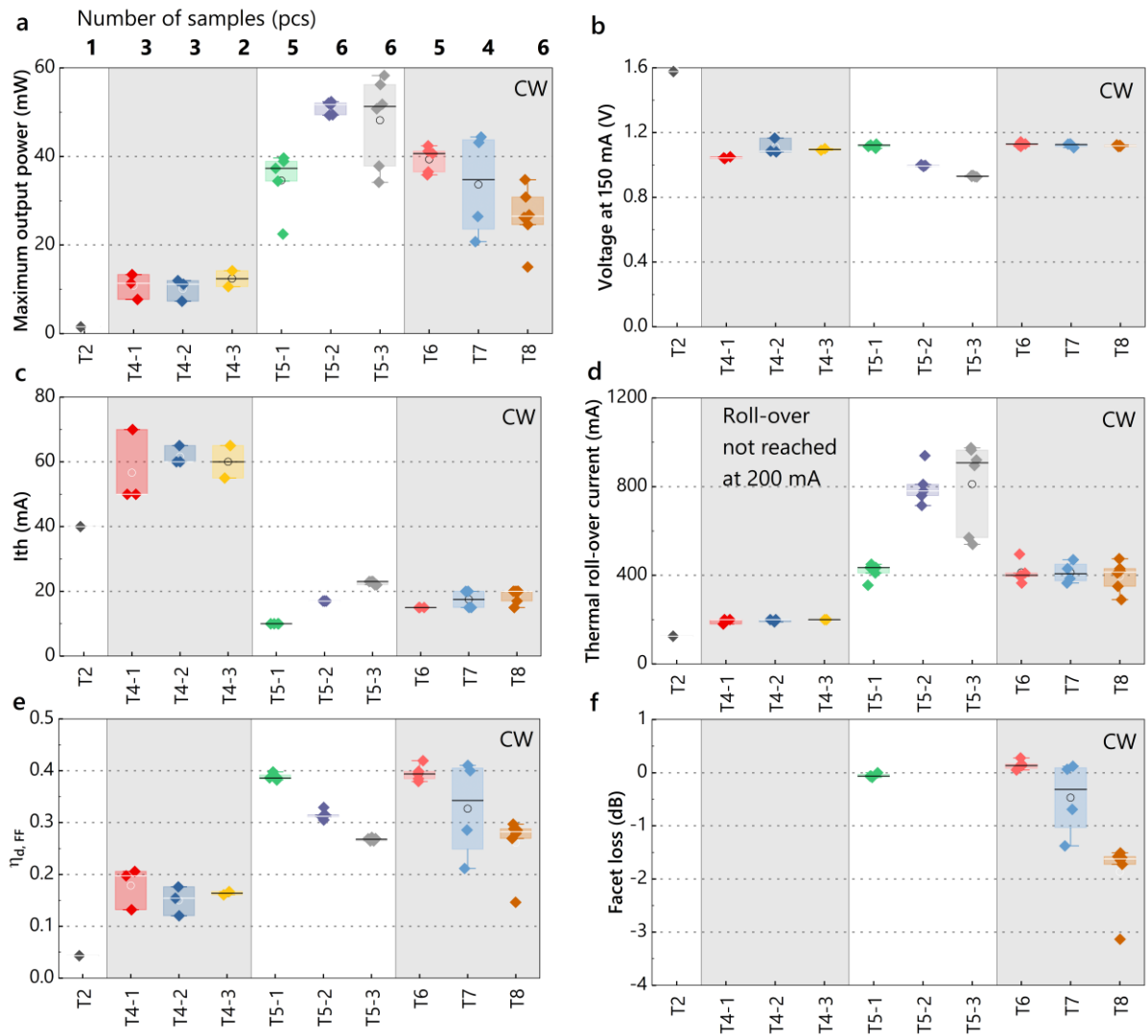

Figure 9. Laser diode device characteristics compared for the samples set. a) maximum measured output power; b) voltage at 150 mA; c) threshold current; d) thermal rollover current; e) front facet differential quantum efficiency, and f) calculated facet loss.

Next, we will analyze the results from the point of view of each test category described in Device design-section. The comparison between the device types is built so that the analysis begins from the simplest possible RWG chip (T5-1) and parameter by parameter adds new comparisons between the device types, finally ending up to a transfer printed RSOA chip. The approach is further elaborated in the Materials and Methods section.

The main conclusions of the analysis, further discussed in the following sections, are:

1. *Etched facet with an optical coating causes no notable penalty to device performance;*
2. *HRC coating material exhibits high absorption losses;*
3. *Epi side n-contacts cause no notable penalty to device performance;*
4. *Wafer level ARC uniformity and deposition control requires process optimization to suppress the lasing at high currents;*
5. *The coupon structure withstands the transfer printing process without noticeable penalty to gain characteristics;*
6. *Transfer printing on Si with BCB glue layer decreases the thermal rollover current.*

## Variation of the facet insertion loss

This section focuses on the analysis of the facet insertion losses. The related devices and their specifications are presented in the Table 1; each row listing the features that are compared across the series. For each row there is a highlighted feature that denotes the difference to the other devices, indicating the parameter change depicted in Figure 4 with the arrows. The color coding at the leftmost column is used across the paper as a visual aid to track the sample types.

Table 1. Device specifications for the RWG test devices used to analyze the material and RWG characteristics and test the etched facets.

**RWG test devices**

| | Type | Epi | Device | Front facet | Back facet | Length | n-contact type | No. of RTAs |
|---|---|---|---|---|---|---|---|---|
| 🟢 | **T5-1** | **2** | **LD** | **RWG cleaved** | **RWG cleaved** | **700 µm** | **n-side** | **x1** |
| 🟣 | T5-2 | 2 | LD | RWG cleaved | RWG cleaved | 1200 µm | n-side | x1 |
| ⚫ | T5-3 | 2 | LD | RWG cleaved | RWG cleaved | 1700 µm | n-side | x1 |
| 🔴 | T6 | 2 | LD | T-bar cleaved | RWG cleaved | 700 µm | n-side | x1 |
| 🔵 | T7 | 2 | LD | T-bar etched LDC | RWG cleaved | 700 µm | n-side | x1 |
| 🟠 | T8 | 2 | LD | T-bar etched LDC | T-bar etched HRC | 700 µm | n-side | x1 |

T5-1 to T5-3 devices, with various cavity lengths of as-cleaved RWGs, were used to characterize the modal gain and internal loss. The data for T5-T8 series also reveals the effects of the etched facets and coatings, referred to as a facet related insertion loss.

By comparing devices T5-1 and T6-T8 we can see that the addition of a front facet T-bar increases the threshold current slightly, indicating a slight increase in the cavity losses. The T-bar waveguide design, cleaved or etched (T6, T7), seems to slightly improve the differential quantum efficiency and the maximum output power. It is only with the addition of the etched back facet and an HRC mirror in T8 that we see a significant performance penalty. The etching of the facet combined with the wafer level ARC and/or HRC seems to increase variation in results. This should be mostly a process optimization problem rather than a critical issue affecting the large -scale manufacturing.

We used the devices (T5-T8) in the facet loss series to analyze the losses by comparing devices with various facet features. The approach is adapted from the laser diode characterization method described by Coldren [51] and is elaborated in the Supplementary information.

The estimated values for facet insertion losses are displayed in Figure 9 f. From these graphs we can see that the facet insertion loss for the best functional devices is similar for the T-bar devices T6 and T7, and the devices T5-1 with the reference facet. The T8 devices with the HRC back facet display largest losses. These losses are assumed to be related to HRC layer thickness deviation from the target structure combined with an unexpectedly high absorption of the mirror materials (this prompts to a change in the materials used for the HRC).

**Variations related to the n-contact**

This section focuses on the analysis of the effects related to epi side n-contact pad. The related devices and their specifications are presented in the Table 2.

For the T4-1 → T4-2 comparison we can see a small penalty in the $\eta_d$, threshold current and voltage when the 2$^{nd}$ RTA treatment is added. Otherwise, the additional treatment does not have a significant effect for TP devices T2 and the test device series T4. From the comparison T4-2 → T4-3 for FP-LDs, which have an additional epi-side n-contact pad, we see no change in the device operation, meaning that the very small epi-side n-pads do not play a significant part in the series resistance.

Table 2. Device specifications for LD and RSOA transfer print coupons without release etch with various mounting schemes and RTA treatments to test the coupon functionality before release etch and the effect of the small n-contact pad.

**Transfer print coupons without release etch**

| Type | Epi | Device | Front facet | Back facet | Length | n-contact type | No. of RTAs |
|---|---|---|---|---|---|---|---|
| T3-1 | 1 | RSOA | T-bar etched ARC | T-bar etched HRC | 700 µm | n-pad | x1 |
| T4-1 | 1 | LD | T-bar etched ARC | T-bar etched HRC | 700 µm | n-pad | x1 |
| T4-2 | 1 | LD | T-bar etched ARC | T-bar etched HRC | 700 µm | n-pad | x2 |
| T4-3 | 1 | LD | T-bar etched ARC | T-bar etched HRC | 700 µm | n-side | x2 |

## Analysis of the transfer print coupons

This section focuses on the analysis of the effects related to the transfer print coupon design and transfer printing. The related devices and their specifications are presented in the Table 3.

Table 3. Device specifications for release etched LD and RSOA transfer print coupons transfer printed on a silicon substrate.

**Transfer print coupons for release etch and transfer printing**

| Type | Epi | Device | Front facet | Back facet | Length | n-contact type | No. of RTAs |
|---|---|---|---|---|---|---|---|
| T1 | 1 | RSOA | T-bar etched ARC | T-bar etched HRC | 700 µm | n-pad | x1 |
| T2 | 1 | LD | T-bar etched ARC | T-bar etched HRC | 700 µm | n-pad | x1 |

T8 used in the facet insertion loss series is the closest in features to T4-1 LD. From the design point of view, these the differences are the lack of epi-side n-contact in T8, the ARC in T4-1, and the different epi materials. This comparison ties the reference devices from T5-T8 series to T2/T4 series. Since the T4-2 → T4-3 comparison does not show a penalty from epi-side n-pad, we assume that the only notable difference for T8 → T4-1 is the ARC, which causes the observed change in the $\eta_d$ and the threshold current. This is because the presence of the ARC affects the lasing in the T4-1 devices and this comparison cannot be used to make relevant conclusion about applicability of the wafer level ARC or the etched facet. The slight drop in voltage is most likely related to different epi wafer as the offset in voltage is visible between T4 and T5-T8 series that have different epi structures.

Finally, the comparison T4-1 → T2 connects the rest of the results to the transfer printed devices providing a reference on the effect the release etch and transfer printing have on the device functionality. Here we see that the transfer printing results in a penalty in the thermal rollover. The total output power and $\eta_d$ for T2 similarly show a clear drop but are not directly comparable with the T4-T8 measurements as the measurements were made using different methods. The voltage experiences a major increase, which is likely connected to different types of current injection probes used (prober needle and wire bonding). The TP coupons have also a lower threshold current than the reference devices.

It is to be noted that for the T2/T4 and T8 devices, with the HRC at the back facet, the differential quantum efficiency for the front facet entails the total output power. This is opposed to the T5-T7 series where the output is divided equally between the front and the back facets. The differential quantum efficiency values in the Figure 9 e are calculated for the front facet. For the T5 series with identical front and back facets, the total $\eta_d$ should be double the front facet value. With T6 and T7 the difference was taken into account when calculating the facet insertion losses by including simulations of the facet reflectance.

**Conclusions**

A GaSb RSOA integrated on a SOI platform using transfer printing was demonstrated for the first time. A methodology and a comprehensive test series was developed to isolate and inspect features inherent to the transfer print process, pointing out future development needs.

The results show that the etched facets of the µ-TP coupons have a high optical quality, avoiding power penalty for lasing operation. The antireflection coating on the front facet applied on wafer level was somewhat off-target as seen in some lasing peaks in the measured spectra indicating need for better means to calibrate coatings deposited on sidewalls of the structures. On the other hand, the deposition of high reflection coating metal mirrors was found to be the largest source of absorption loss due to poor choice of materials and requires changes in materials to avoid losses and consequently increase the output power. Miniaturization of the devices leading to relatively small contact areas and the application of epi-side n-contact pads showed no signs of increase in series resistance. Characterization of the transfer printed devices demonstrated that the epitaxial structure withstands the transfer and produces functional devices with clean LIV curves with an improved threshold current density, but lower thermal roll over current than the reference devices. External cavity laser spectrum showed single mode operation at target wavelength but requires further stabilization of the GaSb coupon for a wider operation window. Strain management was found to be a necessary consideration in the future development to avoid bending of the coupons and thus avoid misalignment with the SOI waveguides.

The demonstrated functionality and methodology proves that the µ-TP can be successfully deployed for GaSb optoelectronics enabling to extend this promising hybrid integration technology for applications requiring integrated light sources with low power consumption emitting beyond 2 µm.

## Materials and Methods

### Methodology to identify performance metrics of the GaSb µTP-devices

As illustrated in Figure 4, we fabricated twelve different test device types alongside the actual transfer print coupons (T1 RSOA and T2 FP-LD) to verify the operation of each element necessary for TP coupon to function. The TP coupon dimensions are 700 µm x 80 µm x 5.4 µm. All fabricated devices have a 3.1 µm wide RWG. The devices can be classified into five categories:

1) Devices with TP coupon geometry (T1, T2) allowing coupon release from III-V substrate;
2) Devices with a geometry similar to T1 and T2 (T3, T4-1), but with modifications to allow testing without release etch;
3) Devices to test the n-contact pad on epi-side of the device (T4-2, T4-3);
4) Devices to test the variability of epi-material;
5) Variants to analyze the facet insertion losses caused by the etched facets (T5-T8).

Device types T1-T4 and T5-T8 were fabricated using separate epi structures, 1 and 2, respectively. Yet, they were included in the same process batch to minimize the influence of the process variations on device performance. The test devices T5-T8 were fabricated on an epi wafer without the release layer structure, but with similar active layers as with the TP epi-structure. For both wafers we used a GaInAlAsSb/GaSb-based type-I QW-gain structure. Some of the test devices have sub-types including variations in the post-processing or packaging for characterization. Figure 4 illustrates the categories and relations between the device variants.

### Device fabrication

The device topography (RWGs, n-contact trench, coupon facets and release layer exposure) was defined using a sequence of UV photolithography steps combined with dielectric or photoresists masks to allow plasma etching (first, the dielectric hard mask etch using reactive ion etch and then the actual semiconductor etch in inductively coupled plasma etcher). All the dielectrics ($SiO_2$ and $SiN_x$) used in the process were deposited using plasma enhanced chemical vapor deposition (PECVD). For contact metals, a lift-off technique with a patterned photoresist mask was used in combination with an oxide removal and electron beam evaporation of contact metals. For n-contact trenches first the actual ohmic contact with semiconductor was created with NiAuGeAu metal stack annealed afterwards in RTA. Later, on top of this, a second contact layer of TiPtAu was deposited together with the p-contact pads through a similar lift off process. After ICP etch, the facets were coated with a dielectric stack in PECVD to form the desired ARC. For the back-facets an additional lift-off lithography step to deposit an Au metal mirror was used. For the final stage, after the exposure of the release layer, the supporting tethers were created using photoresist. After this, the TP coupons were release etched via an immersion in an etchant solution and the ensuing dissolution of the release layer. T1 and T2 were released, and transfer printed. Their electro-optical characterization was only possible after the transfer print. T3 and T4 sub-series were not release etched but were fully processed RSOA and LD coupons with modifications to allow testing without the release etch.

At this point the test samples (T3-T8) were thinned using a lapper to mechanically grind the excess semiconductor from the back side. An n-metal stack of NiAuGeAu was evaporated and annealed in RTA before the devices were diced and mounted for characterization.

**Transfer printing**

We demonstrate two different schemes of transfer printing: 1) a simple transfer on an un-patterned silicon surface, and 2) a transfer on a SOI platform with DBR patterned silicon waveguides to demonstrate an external cavity hybrid laser. To transfer the laser coupons, we used a transparent polydimethylsiloxane (PDMS) stamp dimensioned for the pick-up. The picking and placing of the coupons is controlled by the velocity-dependent adhesion of viscoelastic elastomers [52]. A flat, clean surface location is required for adhesive-less printing by using direct bonding through van der Waals forces [53]. For the initial testing and device characterization, the T1 RSOA and the T2 FP-LD coupons were printed at the edge of a plain Si chip so the light can be edge-coupled into a fiber to enable the characterization.

To demonstrate single mode external cavity laser, a T1 RSOA device was transfer printed on a SOI PIC that had Si waveguides with reflective DBR gratings. The antireflective coating facet of the Si waveguide had an angle of 7° to match the RSOA facet. Figure 5 b shows the SEM image of the transfer printed RSOA coupons with the DBR waveguides. The PIC was cleaved on the waveguide side so that the light could be edge coupled into a single mode lensed fiber for characterization. In these experiments, a SOI PIC with 3 µm thick silicon layer on a 400 nm buried oxide (BOX) layer was used. The waveguide had a rib geometry with a width of 2.6 µm. To engineer transverse matching of the laser and waveguide modes, a trench of 3.4 µm, into the silicon substrate, was etched in front of the waveguide, which forms the printing area for the GaSb gain blocks. The silicon helps in heat dissipation from the coupons due to its excellent thermal conductivity [26].

After first printing tests we measured a gap of 5 µm or less between the RSOA and the waveguide. The gap was formed by the presence of a 3 µm ledge on the III-V coupon edge and the alignment tolerance of manual alignment. Simulations with Lumerical 3D FDTD solver, with PML (perfectly matched layer) boundary conditions, provided an estimated coupling efficiencies of 25% for the 5 µm gap, 38% for the 3 µm gap and 64% for the 0.5 µm gap [17]. By adjusting the coupon design and utilization of automatic alignment, the gap can be reduced to 0.5 µm or less.

To reduce the transition loss from RSOA to the waveguide, the gap was filled with an optical adhesive that is transparent at the 2 µm waveband and cured under UV exposure. The internal stress manifesting after the fabrication resulted in the bending of the coupons that was measured with a white light interferometer; this measurement is shown in Figure 10 revealing a radius of curvature (R) of ~4.5 mm. For better adhesion and to compensate the bending of the coupons, a thin (50 nm) layer of BCB was spin coated on the target surface before printing. The effect can be seen in the SEM images (Figure 5 b) which was taken after the printing. The BCB coating helps to compensate the bend of the coupons, but a thin layer of air gap was still present leading to a vertical misalignment between waveguides.

The reason for the coupon bending is the residual stress of the coupon structure and has been observed also in other release etch works [18,54]. The epi structure could be strain compensated using compound semiconductors engineering, however, we would note that the possible main source of the stress is the dielectric coating stack used as the ARC, covering the entire upper surface of the coupon and the contact metal pads [54]. The PECVD grown dielectrics have internal stress that is linked to the growth parameters and the different thermal expansion coefficients between materials. For e-beam evaporated contact metals, the literature suggests a relation between the film stress and evaporation rate, grain size and layer thickness [55,56]. This means that in the future development the strain design of the µTP coupons is a necessary consideration when designing the epi and dielectric layers to compensate for the stress and to allow planar coupon fabrication. Subsequently, further improvement of the coupon stress will lead to improvement of light coupling to waveguide.

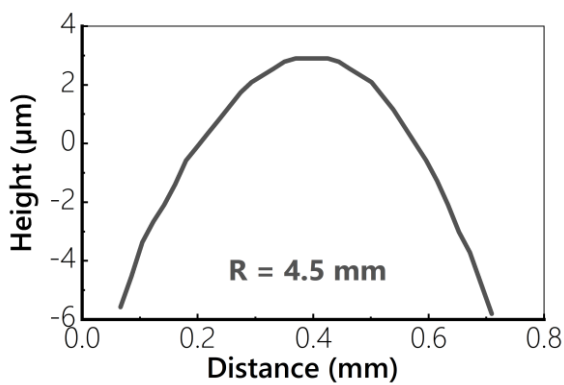

Figure 10. Release etched coupon bending measured with a white light interferometer.

## Supplementary information

In the following section we present the approach we used to extract the insertion losses related to the etched facets. This approach is based on the treatment presented by Coldren [51]. First, we calculated the basic RWG device characteristics: mirror losses $\alpha_m$, internal losses $\langle\alpha_i\rangle$ differential quantum efficiency $\eta_d$, internal quantum efficiency $\eta_i$ and threshold modal gain $\Gamma g_{th}$ using the results from the basic RWG devices with varying cavity lengths (devices T5-1 700 µm, T5-2 1200 µm and T5-3 1700 µm).

Using the threshold condition

$$\Gamma g_{th} = \langle\alpha_i\rangle + \alpha_m \qquad (1)$$

we assume that the internal losses and the internal quantum efficiency remain the same for all test devices (with 700 µm cavity).

The mirror losses, referring to the portion of the mode that escapes the cavity at the output, contributing to the output power, vary from device to device with each device having an unique front and/or back facet reflectance $r_{FF}/r_{BF}$. The mirror losses are calculated using the equation

$$\alpha_m = \frac{1}{L}\ln\frac{1}{r_{FF}r_{BF}} \qquad (2)$$

where L is the cavity length. Each mirror loss type is presented in the Figure 4 a. The reflectance for device T5-1 was calculated using simulated effective refractive index $n_{eff}$ for the used RWG epi structure. For T6-T8 front facets we simulated the $n_{eff}$ value for the T-bar geometry. For T7 and T8 this was used together with a simulated $r_{FF}$ for a dielectric stack imitating the reflection of an uncoated facet. For the back facet of T8 a simulated $r_{BF}$ with a dielectric + Pt/Au coating was used.

To take into account the different conditions of front and back facets and their effect on the differential quantum efficiency, we follow Coldren's treatment to calculated separate values for each facet by introducing the $F_{FF}$ that represents the fraction of the power from front facet relative to the total output power of the device:

$$\eta_{d,FF} = F_{FF}\eta_i \frac{\alpha_m}{\langle\alpha_i\rangle+\alpha_m}. \tag{3}$$

$F_{FF}$ can be calculated with the equation

$$F_{FF} = \frac{t_{FF}^2}{(1-r_{FF}^2)+\frac{r_{FF}}{r_{BF}}(1-r_{BF}^2)}, \tag{4}$$

where $t_{FF}$ is the transmission of front facet. In case the facet is lossy, the condition $t_{FF}^2 + r_{FF}^2 \neq 1$ holds true. We assume that the power lost at the mirror can be detected in the reduction of the differential quantum efficiency measured from the front facet. To find out the effect the addition of the etched facet and the coatings have on the losses we compare these devices (T6-T8) with the reference (T5-1) and extract the difference in the differential quantum efficiencies

$$\Delta\eta_d = \eta_{d,FF,1} - \eta_{d,FF,2}. \tag{5}$$

By combining the Equations (2)-(5) we can formulate an equation for the transmission of the front facet $t_{FF,\,meas}$ for the device under inspection:

$$t_{FF,meas} = \sqrt{\frac{\langle\alpha_i\rangle+\alpha_{m,2}}{\alpha_{m,2}}\left(F_{FF,1}\frac{\alpha_{m,1}}{\Gamma g_{th,1}} - \frac{\Delta\eta_d}{\eta_i}\right)\left((1-r_{FF}^2) + \frac{r_{FF}}{r_{BF}}(1-r_{BF}^2)\right)}. \tag{6}$$

Assuming identical and lossless front and back facets on reference chips T5-1 we set the value $F_{FF,1} = 0.5$. To extract the losses related to the facet we compare the value from Equation (6) to an ideal, lossless mirror with a transmission:

$$t_{FF,ideal} = \sqrt{1-r_{FF}^2}. \tag{7}$$

The facet insertion loss in decibels is then calculated with the equation

$$\alpha_F = 10\log_{10}\left(\frac{t_{FF,meas}}{t_{FF,ideal}}\right). \tag{8}$$